\documentclass[nohyper,twoside]{JHEP3}
\usepackage{graphicx}
\usepackage{epsfig}
\usepackage{float}
\usepackage{amsmath}
\usepackage{amssymb}
\usepackage[square,sort&compress,comma]{natbib}

\newcommand{\be}{\begin{equation}}
\newcommand{\ee}{\end{equation}}
\newcommand{\bea}{\begin{eqnarray}}
\newcommand{\eea}{\end{eqnarray}}
\newcommand{\betahalf}{\frac{\beta}{2}}

\newcommand{\D}{{\cal D}}

\newcommand{\Z}{{\mathcal Z}}

\date{\today}

\title{Geometric Algorithm for Abelian-Gauge Models}

\author{V.~Azcoiti, E.~Follana and A. Vaquero \\ 
Departamento de F\'{\i}sica Te\'orica, Universidad de Zaragoza,\\
Cl. Pedro Cerbuna 12, E-50009 Zaragoza (Spain) \\
E-mail: \email{azcoiti@azcoiti.unizar.es}, \email{efollana@unizar.es}, \email{alexv@unizar.es}} 
\author{G.~Di~Carlo \\ 
INFN, Laboratori Nazionali del Gran Sasso, \\
67010 Assergi,(L'Aquila) (Italy) \\
E-mail: \email{gdicarlo@lngs.infn.it}} 

\abstract{

  Motivated by the sign problem in several systems, we have developed a
  geometric simulation algorithm, based on the strong coupling expansion, which
  can be applied to abelian pure gauge models. We have studied the algorithm
  in the $U(1)$ model in 3 and 4 dimensions, and seen that it is practical and
  is similarly efficient to the standard heat-bath algorithm, but without the
  ergodicity problems which comes from the presence of vortices. We have also
  applied the algorithm to the Ising gauge model at the critical point, and we
  find hints of a better asymptotic behaviour of the autocorrelation time,
  which therefore suggests the possibility of a smaller dynamical critical
  exponent with respect to the standard heat-bath algorithm.}

\begin{document}

\section{Introduction}

The expected scenario for the phase diagram of QCD in the chemical
po\-ten\-tial--tem\-pe\-ra\-tu\-re plane is based on the conjecture that,
besides the well established hadronic and quark-gluon plasma phases, there
exists a new state of matter. This new phase is characteristic of the high
density, low temperature regime: the asymptotic freedom of QCD and the known
instability of large Fermi spheres, in the presence of weak attractive forces,
results in a pairing of quarks with momenta near the Fermi surface (in analogy
with the Cooper pairing in solid state systems at low temperature). A
condensation of quark pairs should be the distinctive signal of the new phase,
and it has indeed been predicted using simplified phenomenological models of
QCD.  Unfortunately the lattice approach, the most powerful tool to perform
first principles, non perturbative studies, is afflicted in the case of finite
density QCD by the well known sign problem that, notwithstanding a large
amount of work by many people \cite{MPLom}, has prevented until now any
significant step towards the understanding of this new phase.

An alternative strategy to overcome the sign problem consists in reformulating
the theory, not in terms of the microscopic degrees of freedom (quarks and
gluons in the QCD case), but in perhaps some other more physical variables
(mesons, baryons, etc. \cite{karsch,Fabrizio}). Following this line, Karsch and
M\"utter \cite{karsch} analyzed the phase diagram of strongly coupled QCD with
SU(3) gauge fields, staggered fermions and four flavors. By in\-te\-gra\-ting
out the SU(3) gauge fields \cite{rossi}, Karsch and M\"utter constructed a new
representation of the QCD partition function as a statistical system of
monomers, dimers and polymers (MDP-system). Even if at finite density some of
the Boltzmann weights were still negative in the new QCD representation, they
claimed that the dominant contributions to the partition function had positive 
weights; and this allowed for the construction of an algorithm that generated
configurations distributed according to the absolute value of the integration
measure, with the sign of the Boltzmann weights being absorbed into the
observables \cite{Aloisio,Fromm}.

Unfortunately the continuum limit of QCD lies not in the strong coupling
regime but in the weak coupling regime, and here the pure gauge Boltzmann
factor strongly changes the gauge integration rules \cite{rossi}, invalidating
the MDP-QCD representation for the partition function. Indeed the MDP
representation would be the zeroth order expansion of the partition function
in powers of $\beta$, the square of the inverse gauge coupling.

The Karsch and M\"utter results for the strong coupling limit of finite
density QCD could, in principle, be extended to higher order in the expansion
in $\beta$. This extension will obviously increase the complexity of the
graphs contributing to the new representation of the QCD partition function,
and furthermore, it is not clear at all whether the sign problem in the new
representation will become a weak sign problem, as in the strong coupling
case, or a severe sign problem \cite{Cherrington}. In any case, the physical
relevance of the system under consideration makes it worthwhile to try such a
line of research. In addition there are other physically relevant gauge models
with a sign problem, such as systems with a topological $\theta$-term in the
action, where these techniques could also be applied
\cite{Meron,Aztheta,RMatrix,Elia,Aztheta2,Entropy,Debbio,AzHaldane,Papa,Topo}.

Several years ago Prokof’ev and Svistunov \cite{pro} proposed new algorithms
for classical statistical spin systems, the worm algorithms, based on the use
of a new representation of the partition function for these systems, which
emerges from the high temperature expansion. The use of this expansion
involves introducing a new configuration space for the Ising model made up
from monomers, dimers and closed paths, with their corresponding exclusion
rules. Prokof’ev and Svistunov showed how these worm algorithms applied to
several spin systems essentially eliminate critical slowing down, with dynamic
critical exponents close to zero, and yet remain local. These results have
been recently corroborated in \cite{Sokal} for the Ising model , reporting a
detailed error analysis and a generalization of the method of \cite{pro}.

Our aim in this paper is to construct and apply algorithms for the strong
coupling representation of the partition function of three and
four-dimensional abelian pure gauge models, in particular $U(1)$ and $Z(2)$
\cite{Chandra}. Independently of our aforementioned motivation, it seems
interesting to us to construct these algorithms and to compare them with
standard ones, such as heat-bath, in order to see if, as reported in
\cite{pro,Sokal,wolff} for spin systems, critical slowing down is also
practically absent in gauge systems. This is precisely the reason why we have
also simulated the three-dimensional Ising-gauge model.  Indeed, contrary to
the $U(1)$ model, the three-dimensional Ising-gauge model, which is dual of
the Ising model in three dimensions, shows a second order phase transition,
were critical slowing down is present in the standard algorithms. The
generalization of the strong coupling representation of the partition function
to the $U(1)$ model with dynamical fermion matter fields, which in principle
suffers from the sign problem (at least in more than two dimensions)
\cite{Wenger,Wolff2}, will be reported in a forthcoming publication.

The paper is organized as follows. In section II we analyze the strong
coupling representation of the partition function, discuss the possible
structures which can appear in the abelian model, and compute the
con\-fi\-gu\-ra\-tion transition probabilities for local changes. Section III
shows how to compute thermodynamic quantities in the new representation, such
as the plaquette energy, specific heat, Wilson loops and correlation
functions. In section IV we describe the implementation of our algorithm,
report numerical results for the three and four dimensional $U(1)$ compact
model, and compare the efficiency of our algorithm against heat-bath.  Section
V contains our results for the three-dimensional Ising-gauge model which, as
stated before, is a good laboratory to check the efficiency of the algorithm
near second order phase transition points. Last section is devoted to
summarize our conclusions and discuss possible extensions of this work.

\section{Abelian groups}

We will consider an (euclidean) abelian $U(1)$ gauge theory on the lattice
with the usual Wilson action,
\be 
S = -\beta {\cal \text{Re}} \sum_{\stackrel{n, \mu, \nu}{\mu < \nu}} 
U_\mu(n) U_\nu(n+\mu) U_\mu^\dagger (n+\nu) U_\nu^\dagger (n) = 
- \sum_{k = 1}^{N_p} \betahalf (U_k + U^*_k) 
\ee 
where $N_p$ is the number of plaquettes of the lattice, $k$ indexes
the plaquettes of the lattice and $U_k$ and $U_k^*$ are the oriented
product of gauge fields along plaquette $k$ and its complex conjugate
respectively.

The starting point of the algorithm is the strong-coupling expansion
of the partition function:
\begin{multline}
\label{Zint}
\Z = \int [\prod_{n, \mu}dU_\mu(n)] \prod_{k = 1}^{Np} 
e^{\betahalf (U_k + U^*_k)} =  
\int [\D U] \prod_{k} \left\{ \sum_{n = 0}^{\infty}
  \frac{(\betahalf)^n}{n!} 
(U_k + U^*_k)^n \right\} = \\
\int [\D U] \prod_{k} \left\{\sum_{n = 0}^{\infty} \frac{(\betahalf)^n}{n!} 
\left[\sum_{j = 0}^{n} \binom{n}{j} (U_k)^j (U_k^*)^{n-j}\right] \right\} = \\
\int [\D U] \prod_{k} \left\{\sum_{j_1,j_2 = 0}^{\infty} 
\frac{(\betahalf)^{j_1+j_2}}{(j_1+j_2)!} \binom{j_1+j_2}{j_1} (U_k)^{j_1} (U_k^*)^{j_2} \right\}
\end{multline}
If we expand the product over the plaquettes, we get a sum of terms, which can
be uniquely labeled by giving a couple of natural numbers for each plaquette
$(n_k^\alpha, \bar n_k^\alpha)$, where $n_k^\alpha$ and $\bar n_k^\alpha$
correspond to the powers of $U_p$ and $U_p^*$, and the super-index $\alpha$
refers to a specific term in the expansion.  After integration over the gauge
fields, the only non-vanishing contributions to $\Z$ are those in which every
link corresponding to a plaquette with $(n_k^\alpha, \bar n_k^\alpha) \ne
(0,0)$ appears accompanied by its complex conjugate. Each of these
contributions therefore can consist only of plaquettes with $n_k^\alpha = \bar
n_k^\alpha$, and plaquettes which form closed orientable surfaces. We can
write
\be
\label{Z}
\Z(\beta) = \sum_{\alpha} C^\alpha(\beta)
\ee
where $\alpha$ indexes the set of non-vanishing contributions. The
(non-negative) weight corresponding to such a contribution is given by
\be
\label{C}
C^\alpha(\beta) = \prod_{k} \frac{\beta^{n_k^\alpha + \bar n_k^\alpha }}
{2^{n_k^\alpha + \bar n_k^\alpha}
  n_k^\alpha! \bar n_k^\alpha !} 
\ee
Now we can view $\Z$ as the partition function of a new system, where a
configuration $\alpha$ is characterized by the set of numbers $(n_k^\alpha,
\bar n_k^\alpha)$ corresponding to a non-vanishing contribution to the strong
coupling expansion, with probability $w^\alpha = C^\alpha / \Z$, and implement
a Monte Carlo in this new configuration space. We define elementary Monte
Carlo moves consisting in adding a ``double plaquette'', $(n_k^\alpha, \bar
n_k^\alpha) \rightarrow (n_k^\alpha +1, \bar n_k^\alpha +1)$ (as well as the
opposite move, if both $n_k^\alpha, \bar n_k^\alpha > 0$), and adding an
elementary cube (as well as removing it if present). This ensures that we stay
inside the set of non-vanishing contributions. We choose the transition
probabilities to be proportional to the weights of the corresponding
configurations in the partition function, which ensures detailed balance. This
can be done efficiently, and it is also easy to see that the algorithm is
ergodic.
\footnote{A single surface that wraps around the lattice cannot be
  created by this algorithm starting from the empty configuration, but
  this is only a finite-volume correction.}

The gauge $Z_p$ model can be handled in the same way, the only
difference being that now there are additional non-vanishing
contributions given by powers of $p$ of the plaquette or its complex
conjugate, that is, we must allow contributions with \mbox{$n_k^\alpha
  - \bar n_k^\alpha = 0 \pmod p$.} The weight of a configuration is
given by the same expression as before \eqref{C}. The only
modification required in the algorithm is adding a new move which
creates or destroys a $p$ power of the plaquette or its complex
conjugate.

\section{Observables}

The computation of observables is quite easy in this representation. The
definition of the observable plaquette\footnote{We must remark that the
  observable plaquette and the plaquettes living in our lattice are not the
  same, although they are strongly related. The observable plaquette refers to
  the minimal Wilson loop, whereas the plaquettes refers to geometric
  entities, living on the lattice. In order to keep the discussion clear, we
  will always refer to the minimal Wilson loop as \emph{\underline{observable}
    plaquette}.} is
\begin{equation}
\langle P_{\Box}\rangle = \frac{1}{N_P}\partial_{\beta}
\ln\left(\Z\right)
\label{MeanPlq}
\end{equation}
Using \eqref{Z} and \eqref{C}, we obtain
\begin{equation}
\langle P_{\Box}\rangle = \frac{1}{N_P}\partial_{\beta}
\ln\left(\sum_{\alpha \in \mathcal{C}} C^\alpha\right) =
\frac{1}{N_P}\frac{1}{\Z}\sum_{\alpha \in \mathcal{C}}\frac{\left(
n^\alpha+\bar n^\alpha\right)}{\beta} C^\alpha
\label{MeanPlq2}
\end{equation}
with 
$$n^\alpha=\sum_{k=1}^{N_P}n^\alpha_k$$
$$\bar n^\alpha=\sum_{k=1}^{N_P}\bar n^\alpha_k$$
The quantity $C^\alpha/\Z$ is the probability $w^\alpha$ of each
configuration $\alpha$. The mean of the observable plaquette is then
\begin{equation}
\langle P_{\Box}\rangle = \frac{1}{\beta N_P}\sum_{\alpha \in \mathcal{C}}
w^\alpha \left(n^\alpha+\bar n^\alpha\right)
\label{MeanPlq3}
\end{equation}
that is to say, the observable plaquette is equal to the mean value of the sum
of the occupation numbers $n^\alpha$ plus $\bar n^\alpha$, divided by $\beta$
and normalized by $N_p$.

Another interesting observable is the specific heat:
\begin{equation}
C_V = \partial_{\beta} \langle P_\Box\rangle
\label{CVDef}
\end{equation}
We can profit from the previous expression of $P_\Box$
\eqref{MeanPlq3} to find the following equation
\begin{equation}
C_V = \frac{1}{N_P} \left\{\sum_{\alpha \in\mathcal{C}}
w^\alpha \frac{\left( n^\alpha+\bar n^\alpha\right)^2}{\beta^2}
-\left(\sum_{\alpha \in \mathcal{C}} w^\alpha 
\frac{\left( n^\alpha+\bar n^\alpha\right)}{\beta}\right)^2
-\sum_{\alpha \in \mathcal{C}} w^\alpha 
\frac{\left( n^\alpha+\bar n^\alpha\right)}{\beta^2} \right \}
\label{CVDev2}
\end{equation}

Sometimes, it is interesting to compute correlation observables, such as the
Wilson loop (larger than the single observable plaquette), or the
plaquette--plaquette correlation function. To compute these it will prove
helpful to introduce a pair of variable coupling constants $\{\beta_k, \bar
\beta_k\}$, which depend on the plaquette site $k$, in such a way that the
partition function reads now
\begin{equation}
\Z(\beta_j, \bar\beta_j) = \int\left[dU\right] \prod_k
e^{\left( \frac{\beta_k}{2}U_k 
+ \frac{\bar\beta_k}{2}U_k^\star\right)}
\label{ZNoInt2}
\end{equation}
The weight of the configurations changes accordingly
\begin{equation}
C^\alpha(\beta_j, \bar \beta_j) = \prod_{k=1}^{N_P}
\frac{\beta_k^{n^\alpha_k}\bar\beta_k^{\bar
n^\alpha_k}}{2^{n^\alpha_k+\bar n^\alpha_k} \left(n^\alpha_k
!\right)\left(\bar n^\alpha_k !\right)}
\label{WilsonAlpha}
\end{equation}
Now the correlation functions or the Wilson loops are computed by
simple derivation, and then taking all the $\beta_k$, $\bar\beta_k$ to
the same value. For instance, the $2\times 1$ Wilson loop can be
calculated as
\begin{equation}
\langle P_{W_{2\times 1}}\rangle =
2^2\lim_{\beta_j,\bar\beta_j\rightarrow\beta} 
\frac{\sum_{\alpha \in \mathcal{C}}
\partial_{\beta_k}\partial_{\beta_{k+1}}
C^\alpha\left(\beta_j,\bar\beta_j\right)}
{\sum_{\alpha \in \mathcal{C}}C^\alpha\left(\beta_j,\bar\beta_j\right)}
=2^2\sum_{\alpha \in \mathcal{C}}\frac{
n^\alpha_k
n^\alpha_{k+1}}{\beta^2} w^\alpha
\label{MeanW2}
\end{equation}
where $k$ and $k+1$ are contiguous plaquette sites. The generalization of this
result to larger planar Wilson loops is straightforward. Indeed, the
expectation value of any planar Wilson loop can be computed as the mean value
of the product, for all plaquettes enclosed by the loop, of the occupation
number of plaquettes in each plaquette site, multiplied by a factor $2/\beta$
to a power which is just the number of plaquettes enclosed by the loop. This
observable is quite remarkable, for it is computed as a product of occupation
numbers, and features a $(2/\beta)^{Area}$ factor, which eventually may become
exponentially large or small as the size of the loop increases. All these
particular facts render this observable hard to compute, as we will see in the
numerical results.

In the same way we can also obtain the correlation functions for two
arbitrary plaquettes on the lattice:
\begin{equation}
\label{uu}
\langle U_k U_l\rangle =
2^2\lim_{\beta_j,\bar\beta_j\rightarrow\beta} 
\frac{\sum_{\alpha \in \mathcal{C}}
\partial_{\beta_k}\partial_{\beta_{l}}
C^\alpha\left(\beta_j,\bar\beta_j\right)}
{\sum_{\alpha \in \mathcal{C}}C^\alpha\left(\beta_j,\bar\beta_j\right)}
=2^2\sum_{\alpha \in \mathcal{C}}\frac{
n^\alpha_k
n^\alpha_l}{\beta^2} w^\alpha
\end{equation}
and
\begin{equation}
\label{uudag}
\langle U_k U^*_l\rangle =
2^2\lim_{\beta_j,\bar\beta_j\rightarrow\beta} 
\frac{\sum_{\alpha \in \mathcal{C}}
\partial_{\beta_k}\partial_{\bar\beta_{l}}
C^\alpha\left(\beta_j,\bar\beta_j\right)}
{\sum_{\alpha \in \mathcal{C}}C^\alpha\left(\beta_j,\bar\beta_j\right)}
=2^2\sum_{\alpha \in \mathcal{C}}\frac{
n^\alpha_k
\bar n^\alpha_l}{\beta^2} w^\alpha
\end{equation}
Finally from \eqref{uu} and \eqref{uudag}, and taking into account the
symmetry of the model, we can write:
\bea
\label{uuR}
\left<{\cal\text{Re}}U_k {\cal\text{Re}} U_l\right>_c &=& 
\frac{1}{\beta^2}\left< \left(n_k + \bar n_k \right) 
\left(n_l + \bar n_l \right) \right>
- \left< n_k+\bar n_k \right> 
\left< n_l + \bar n_l \right>  \\
\label{uuI}
\left<{\cal\text{Im}}U_k {\cal\text{Im}} U_l\right>_c &=& 
- \frac{1}{\beta^2}\left< \left(n_k - \bar n_k \right) 
\left(n_l - \bar n_l \right) \right>
\eea
where the brackets denote average over configurations.

\section{Implementation and numerical results}

In order to see our algorithm (which we should call, from now on,
\emph{geometric algorithm}) at work, we have performed numerical simulations
of several lattice gauge theory systems in three and four dimensions.  Our
aim is to check the goodness of our approach, comparing the results we obtain
using the geometric algorithm with those coming from more standard ones;
hence we want to compare the properties of the algorithm itself, in terms of
numerical efficiency and autocorrelation times with, for example, the usual
heat-bath algorithm. We have in mind the results of \cite{Sokal}, where it was
claimed that, with a similar algorithm, at a second order critical point in a
non-gauge system, there is a very strong reduction in the critical slowing
down.

Let us start with the three dimensional U(1) lattice gauge model: this model
is known to have a single phase from strong to weak coupling.  We have chosen
to measure two simple observables, namely the plaquette observable and the
specific heat, following the definitions given in the preceeding Section. We
have simulated the model with our algorithm and with a standard heat-bath for
a large interval of $\beta$ values using a $12^3$ lattice; we allowed the
system to reach thermalization for $5 \times 10^4$ iterations and then
measured the observables for $10^6$ iterations. Errors were evaluated using a
Jackknife procedure. The results are shown in Fig \ref{U1-3D}.
\FIGURE[float]{
\resizebox{12 cm}{!}{\includegraphics[angle=-90]{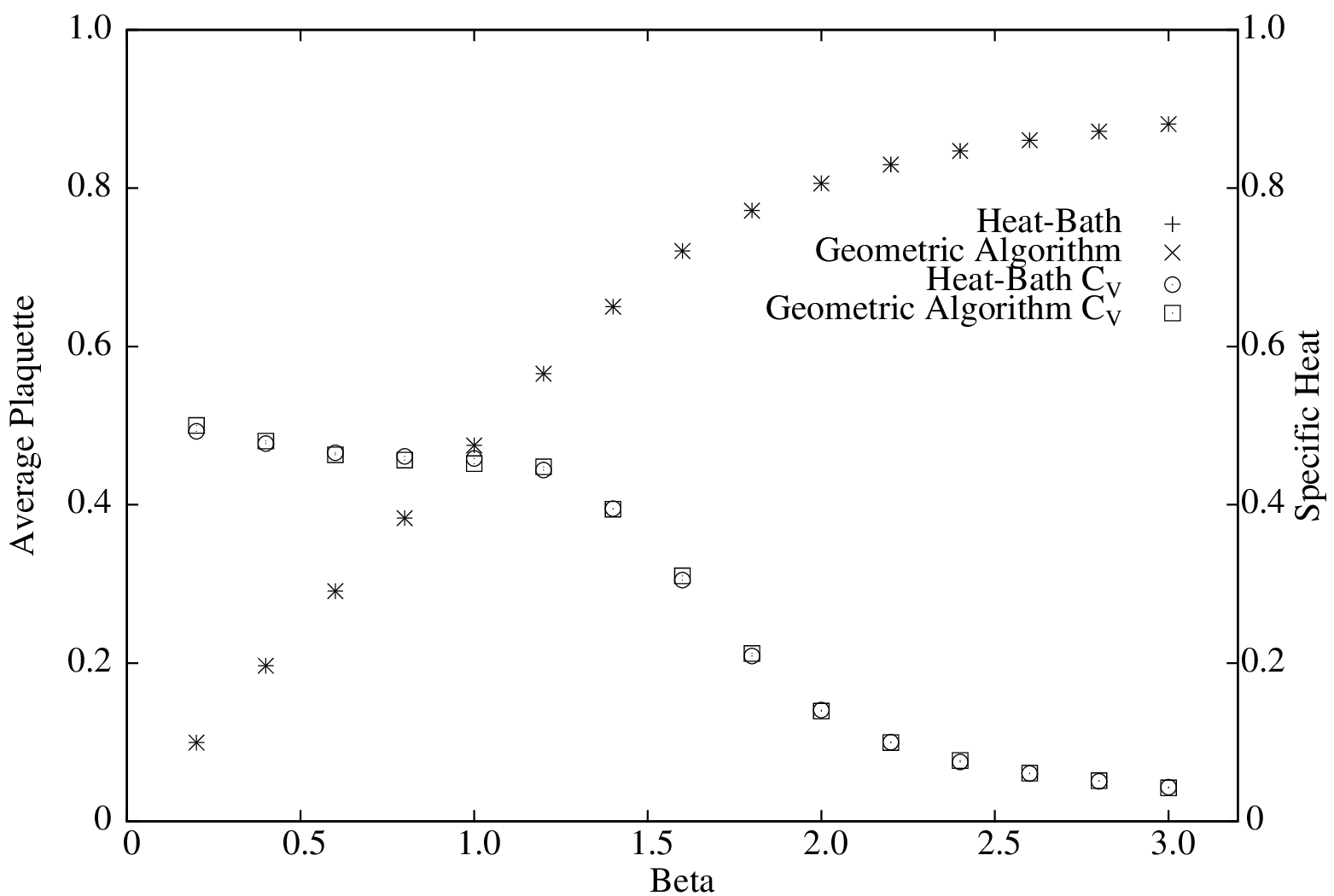}}
\caption{Three dimensional U(1) lattice gauge system. 
Errors are smaller than symbols.}
\label{U1-3D}}
We can easily see in this figure that the two simulations give
essentially the same results.

Almost the same situation can be depicted also for the four
dimensional U(1) model; the results of a similar set of simulations,
performed with the two algorithms on a $16^4$ lattice, are shown in
Fig \ref{U1-4D}.
\FIGURE[float]{
\resizebox{12 cm}{!}{\includegraphics[angle=-90]{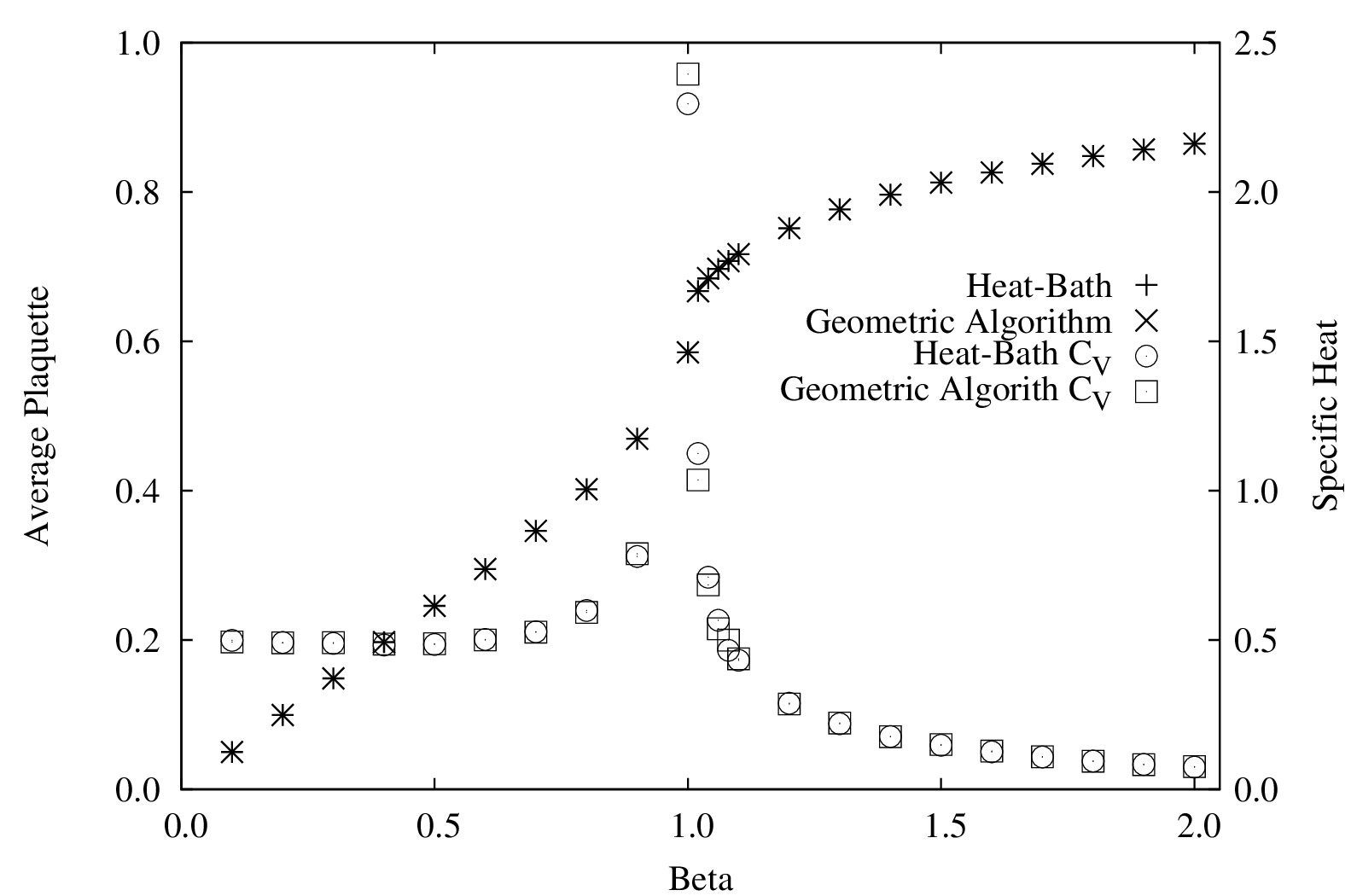}}
\caption{Four dimensional U(1) lattice gauge system; please note the different
  scale for the specific heat on the right.  Largest errors (those on the
  pseudo-critical point) are smaller than symbols.}
\label{U1-4D}}
Here the only difference can be seen near the phase transition point.
Remember that due to the difference in finite volume terms between the two
algorithms, the precise pseudo-critical coupling value at finite volume has to
be slightly different. We have calculated a few more points very close to the
critical beta for each algorithm, and the results can be seen in tables 1 and
2. Note the much larger value of the specific heat for some of these points as
compared with fig.\ref{U1-4D}.
\DOUBLETABLE{
\begin{tabular}{ccc}
$\beta$ & Plaquette & $C_v$ \\
\hline \\
1.010700 &   0.62545(34)  &    24.1(3) \\
1.010850 &   0.64014(143) &    88.2(2) \\
1.010900 &   0.64940(54)  &    47.9(4) \\
1.011100 &   0.65473(6)   &    2.6034(1) \\
1.011600 &   0.65576(12)  &    2.5559(2) \\
\end{tabular}}
{\begin{tabular}{ccc}
$\beta$ & Plaquette & $C_v$ \\
\hline \\
1.011120  &  0.6373(45)  &  67.2(7) \\
1.011160  &  0.6423(46)  &  67.4(3.6)   \\
1.011420  &  0.6549(1)   &  2.9660(2) \\
\end{tabular}
}{Heat-bath results near $\beta_c$.}{Geometric results near $\beta_c$.}

These results allow us to infer that the geometric approach is able to
reproduce all the features of the models under investigation, and when
differences are seen, they can be easily explained on the difference between
finite volume terms. To study more carefully these differences we have
calculated $\beta_c(L)$, the critical coupling for each algorithm at different
lattice sizes. We present in table~\ref{betac} the results. We also include in
the table the results of a fit of $\beta_c(L)$ for the three largest lattices
from each set to the finite-size scaling law expected for a first-order phase
transition \cite{Azcoiti}. This gives a good fit and consistent results for
the infinite volume limit.
\TABLE{
   \caption{$\beta_c(L)$ for heat-bath and geometric algorithm respectively.}
   \label{betac}
\begin{tabular}{ccc}
$L$ & $\beta_c(L)$ (heat-bath) & $\beta_c(L)$ (geometric) \\
\hline \\
6   &    1.00171(8)  &        1.00878(20) \\
10  &    1.00936(11)  &       1.01062(7)  \\
12  &    1.01027(8)    &      1.01100(7) \\
14  &    1.01064(4)     &     1.01103(20) \\
16  &    1.01084(17)     &    1.01116(14) \\
$\infty$ & 1.01108(10)     &  1.01120(22) \\
\end{tabular}
}

The presence of two clearly different phases in this model, namely a confining
and a Coulomb one, allows us to study the behaviour of the Wilson loop results
in two different physical situations; as above, we have also performed standard
simulations for a cross check between the two approaches.  In
Figs. \ref{wlooparea} and \ref{wloopperimeter} we report the behaviour of the
Wilson loop in both phases (confining in Fig. \ref{wlooparea} and Coulomb in
Fig. \ref{wloopperimeter}) and in lattices of different size ($12^4$, $14^4$,
$16^4$).
\FIGURE[float]{
\resizebox{12 cm}{!}{\includegraphics[angle=-90]{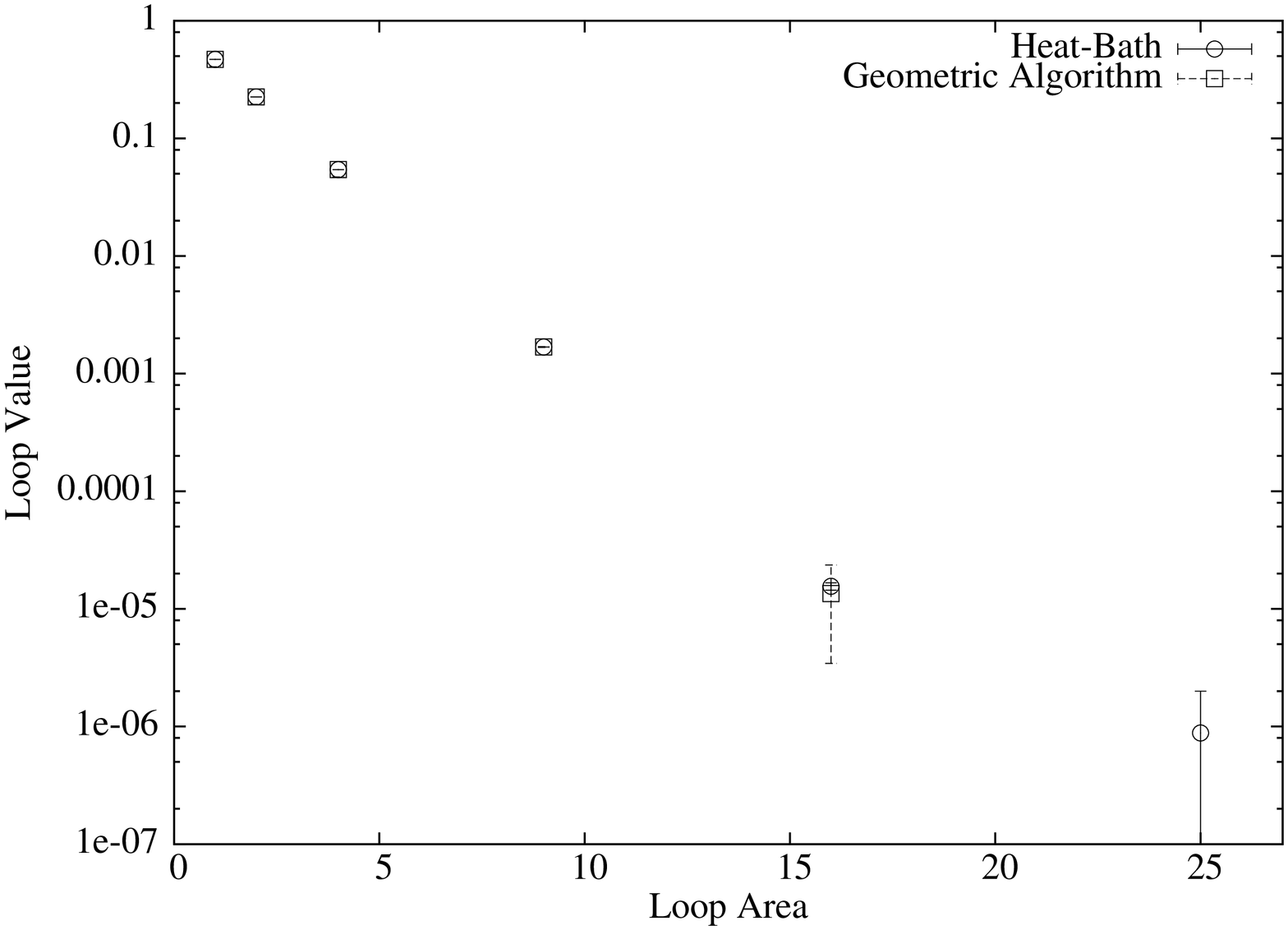}}
\caption{Real part of the Wilson loop versus the loop area for the confining
  phase ($\beta=0.9$) in the four-dimensional U(1) gauge model. Notice the
  absence of the $5\times5$ loop in the geometric algorithm. The lattice
  volume was $16^4$.}
\label{wlooparea}}
\FIGURE[float]{
\resizebox{12 cm}{!}{\includegraphics[angle=-90]{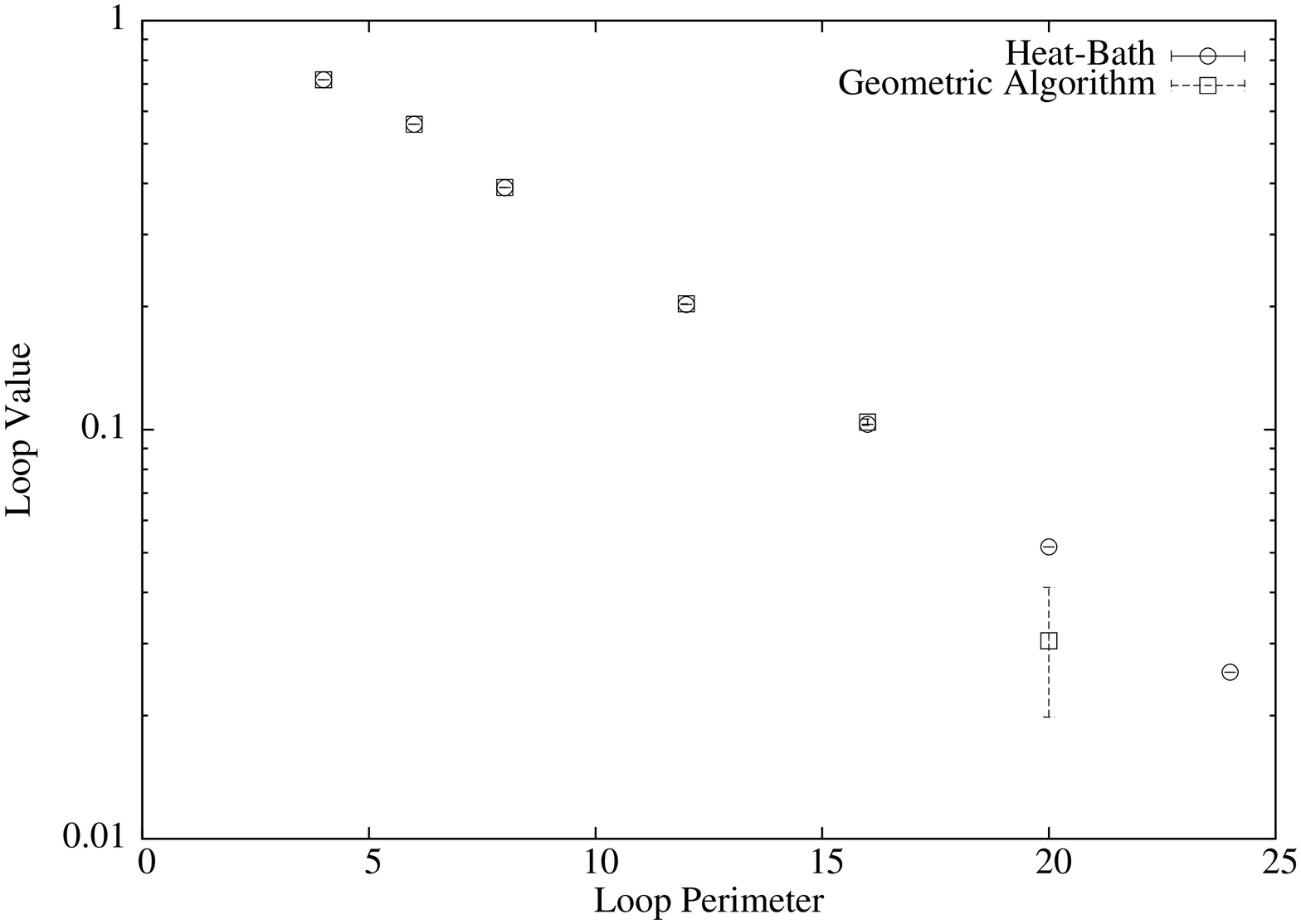}}
\caption{Real part of Wilson loop versus the loop perimeter for the Coulomb
  phase ($\beta=1.1$) in the four-dimensional U(1) gauge model. The lattice
  volume was $16^4$.}
\label{wloopperimeter}}

These figures deserve some comments. The geometric algorithm seems to
suffer from larger statistical errors than the heat-bath method,
regardless of the phase of the system. To understand this result, we
should have a close look at the inner machinery of the algorithm, in
particular, the way the Wilson loop is computed (see
eq. \eqref{MeanW2}). First of all, the mean value of the Wilson loop
is computed as a sum of integer products, implying the existence of
large fluctuations between configurations.  For example, doubling the
occupation number of a single plaquette doubles the value of the
loop. This is a quite common fluctuation at the $\beta$ values of our
simulations, and the fluctuations will increase as the loop (and
therefore the number of plaquettes) grows. To complicate the
computation further, we are trying to calculate an exponentially small
quantity by summing integer numbers. The discrete nature of this
computation tells us that non-zero values of the quantity must appear
with an exponentially small probability. This explains the inherent
difficulties of the large Wilson loops ($4\times4$ and greater)
measurement in the confining phase. The result is shown in
Fig. \ref{wlooparea}: the mean value of the $5\times5$ Wilson loop was
exactly zero in the geometric algorithm, which is of course
wrong. Finally, the expectation value of the Wilson loop is
proportional to a $\left(2/\beta\right)^{A}$ factor, with $A$ the
loop area. This value may become huge (or tiny) for large loops and
low (or high) values of beta, thus enhancing the problems that arise
from the discreteness of the algorithm~\footnote{See
  \cite{Panero1,Panero2} for a discussion of some numerically
  efficient algorithms for the calculation of large Wilson loops and
  Polyakov loop correlators.}.

\FIGURE[float]{
\resizebox{12 cm}{!}{\includegraphics[angle=-90]{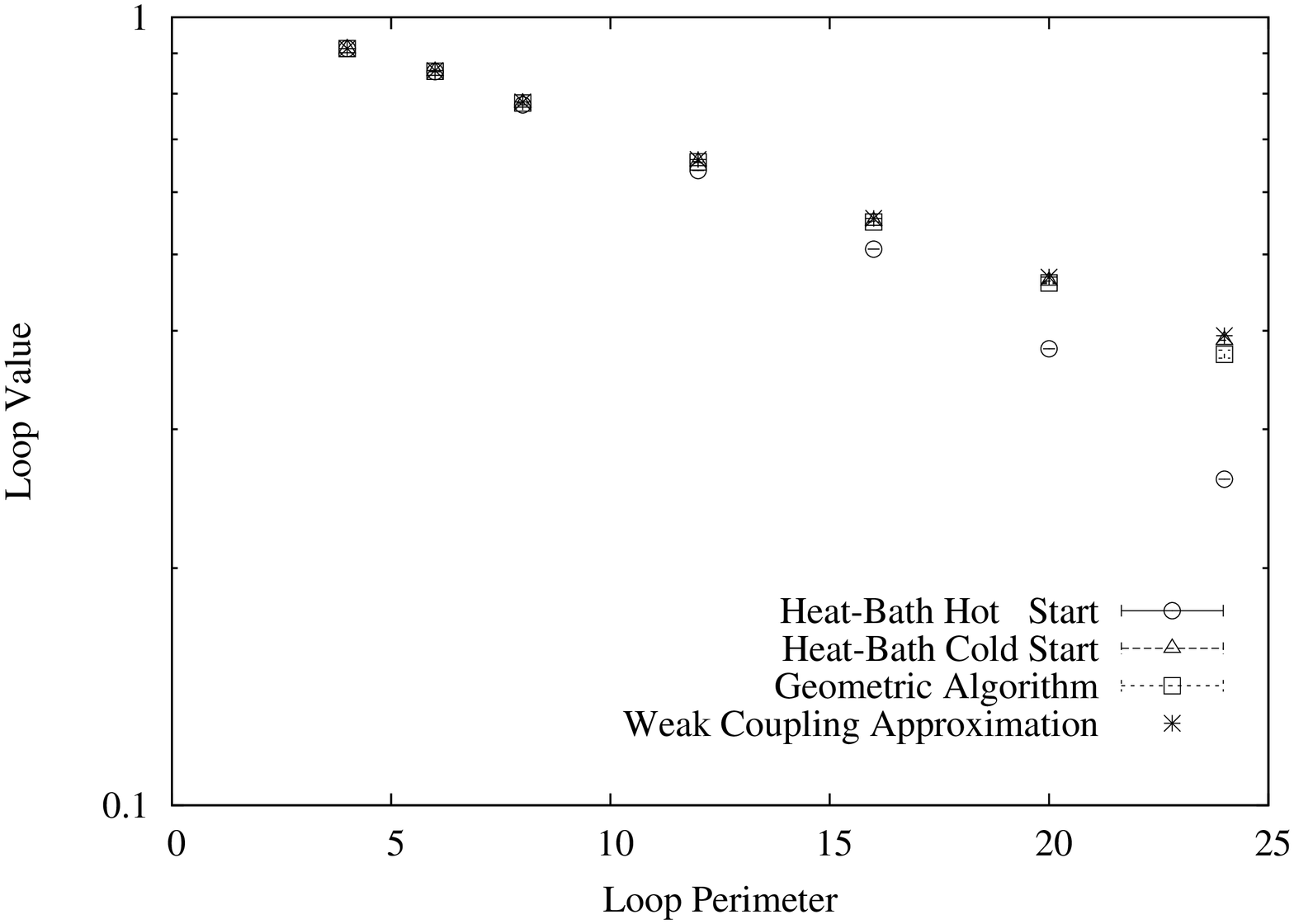}}
\caption{Real part of Wilson loop versus the loop perimeter for a large
  $\beta$ value ($\beta=3.0$) in the four-dimensional U(1) gauge model. Notice
  the difference in performance between the hot and the cold starts of the
  heat-bath algorithm. The lattice volume was $12^4$.}
\label{wloop3}}

Notwithstanding the stronger fluctuations in the large Wilson loops within the
geometric algorithm discussed above, it has a clear advantage
against heat-bath: it does not suffer from ergodicity problems. Indeed the
results for the Wilson loop at $\beta=3$ reported in Fig. \ref{wloop3}
strongly support the previous statement. The points obtained with the
geometric algorithm nicely follow the weak coupling prediction of \cite{weak},
whereas the heat-bath results for large Wilson loops, obtained from a hot
start, clearly deviate from the analytical weak coupling prediction. The
origin of this anomalous behavior in the heat-bath case is related
to the formation of vortices, which are metastable states, that become
extremely long lived in the Coulomb phase \cite{monos}.

We have also calculated the plaquette--plaquette correlation (of the real
\eqref{uuR} and of the imaginary \eqref{uuI} parts) in both phases and for
plaquettes living in the same plane. Here we expect a much milder behaviour
for the geometrical algorithm, for there is no large
$\left(2/\beta\right)^{A}$ factor, and the fluctuations are reduced to a
couple of plaquettes. The results are shown in Figs. \ref{Corrb09},
\ref{Corrb11}, \ref{ICorrb09} and \ref{ICorrb11}.

\FIGURE[float]{
\resizebox{12 cm}{!}{\includegraphics[angle=-90]{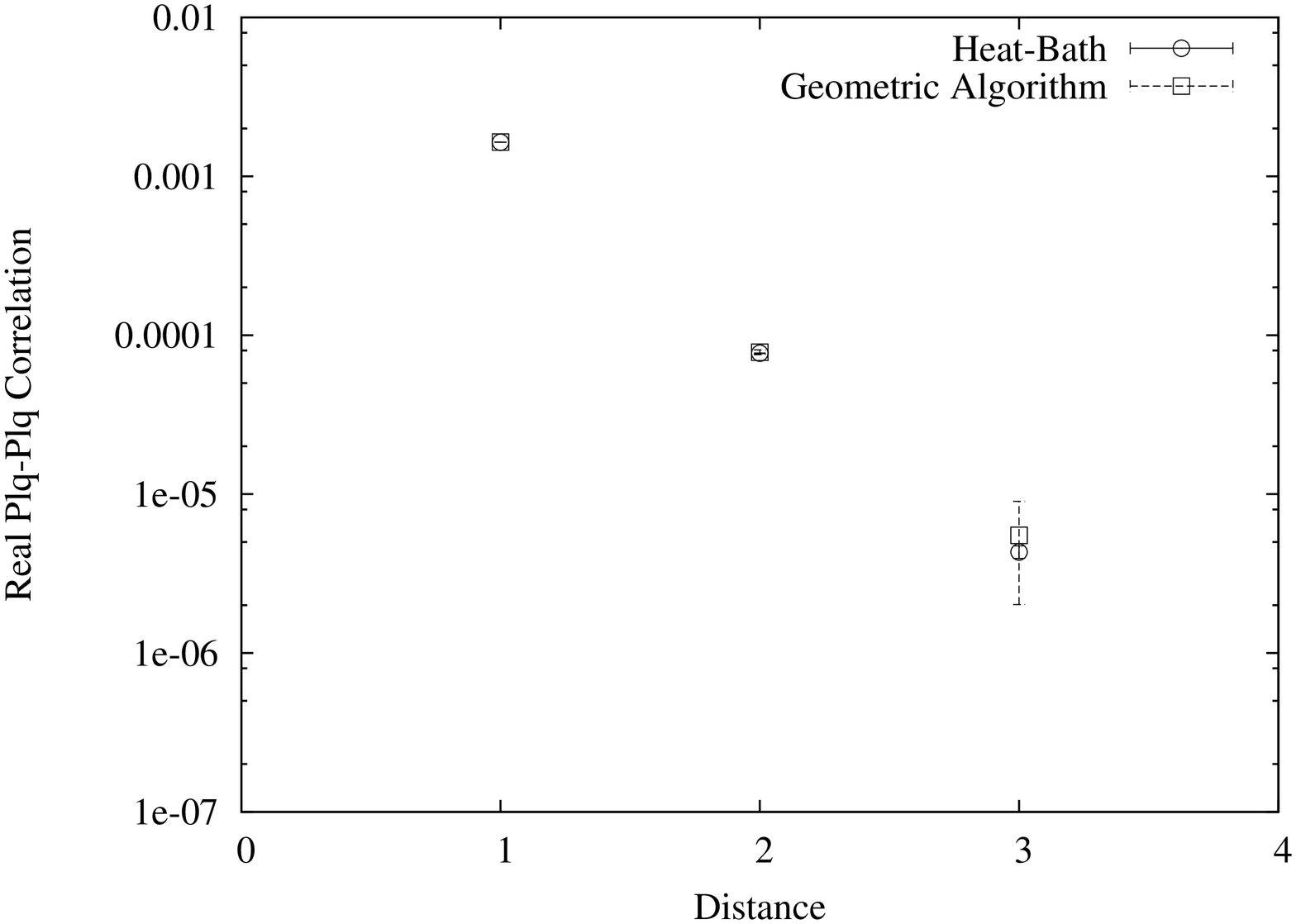}}
\caption{Correlation function of the real part of the plaquette versus
  plaquette--plaquette distance in lattice units, for the four-dimensional
  U(1) lattice gauge model in the confining phase ($\beta = 0.9$). Beyond
  distance $4$, the error became far larger than the expectation value of the
  correlation. The lattice volume was $16^4$.}
\label{Corrb09}}
\FIGURE[float]{
\resizebox{12 cm}{!}{\includegraphics[angle=-90]{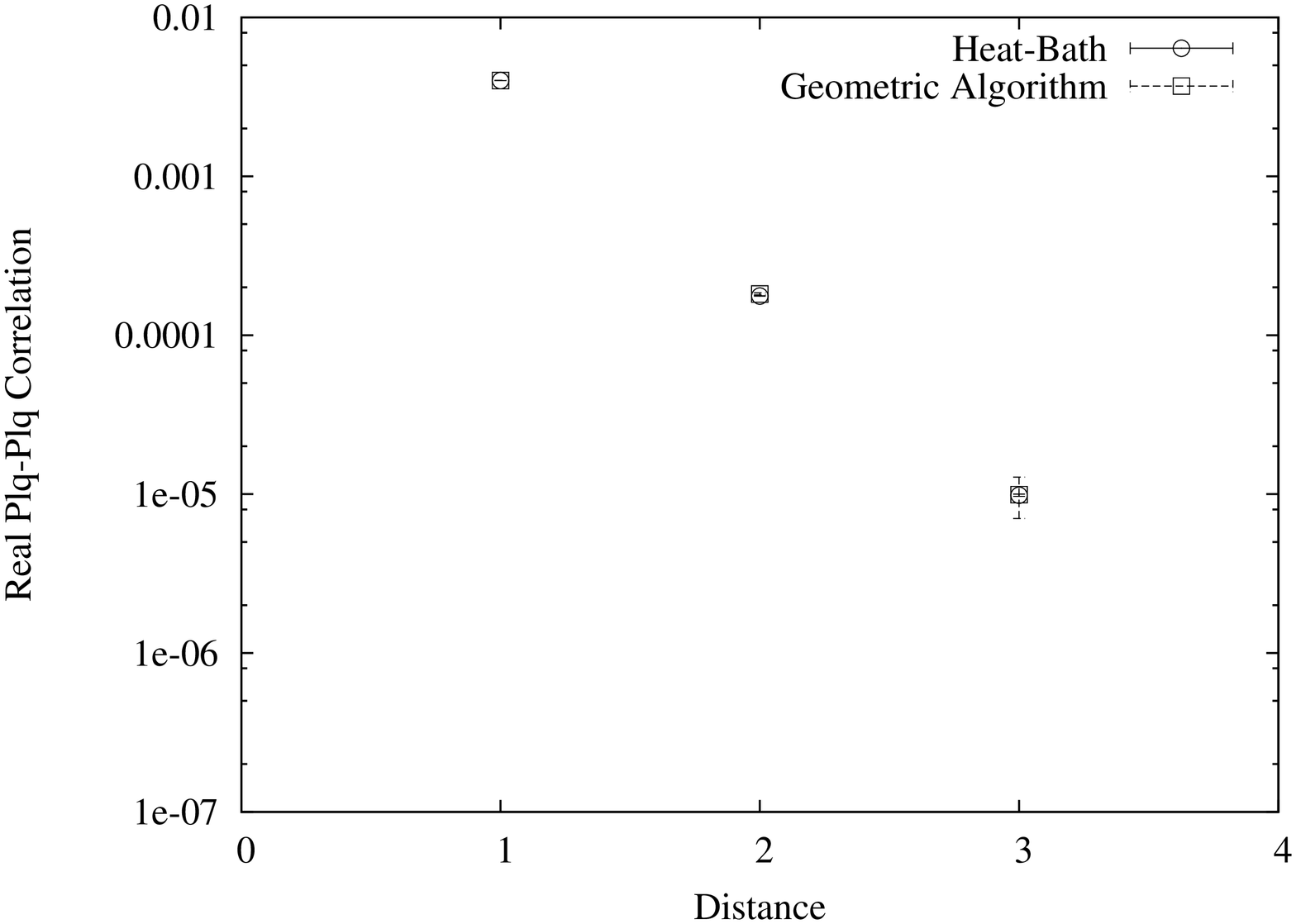}}
\caption{Correlation function of the real part of the plaquette versus
  plaquette--plaquette distance in lattice units, for the four-dimensional
  U(1) lattice gauge model in the Coulomb phase ($\beta = 1.1$). Beyond
  distance $4$, the error became far larger than the expectation value of the
  correlation. The lattice volume was $16^4$.}
\label{Corrb11}}
\FIGURE[float]{ \resizebox{12 cm}{!}{\includegraphics[angle=-90]{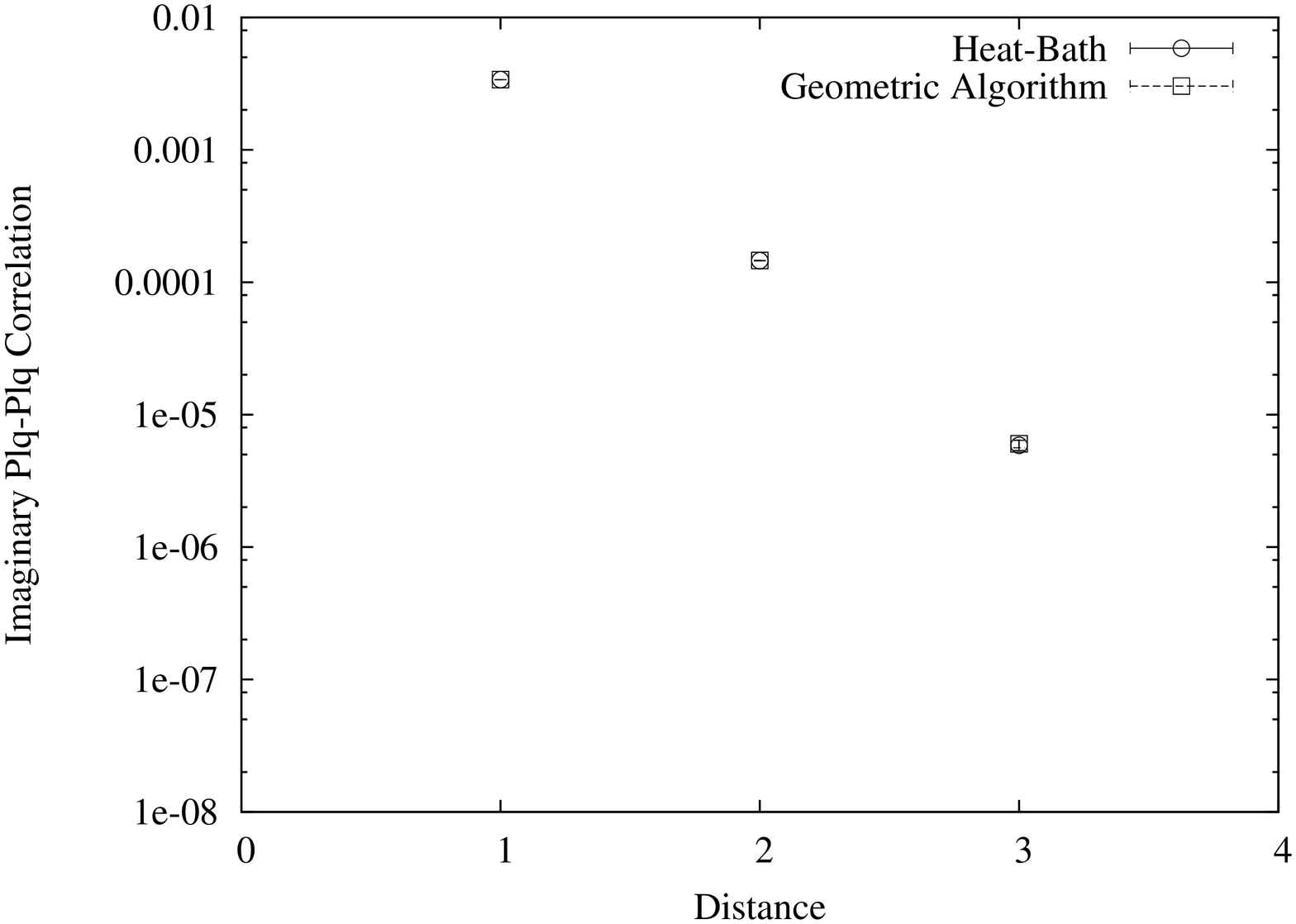}}
\caption{Correlation function of the imaginary part of the plaquette versus
  plaquette--plaquette distance in lattice units, for the four-dimensional
  U(1) lattice gauge model in the confining phase ($\beta = 0.9$). Beyond
  distance $4$, the error became far larger than the expectation value of the
  correlation. The lattice volume was $16^4$.}
\label{ICorrb09}}
\FIGURE[float]{
\resizebox{12 cm}{!}{\includegraphics[angle=-90]{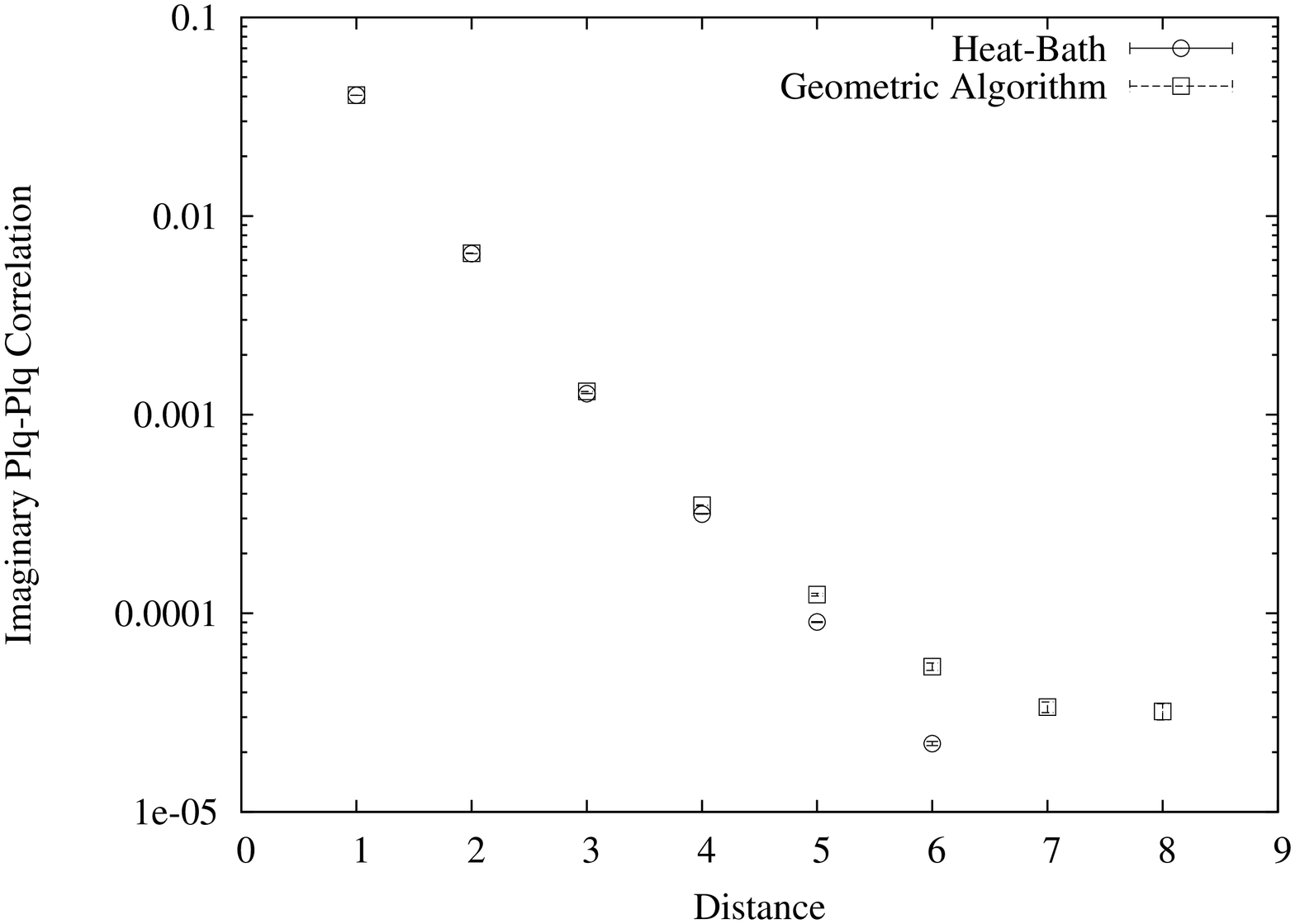}}
\caption{Correlation function of the imaginary part of the plaquette versus
  plaquette--plaquette distance in lattice units, for the four-dimensional
  U(1) lattice gauge model in the Coulomb phase ($\beta = 1.1$). Notice the
  different behaviour of the algorithms at large distances. The lattice volume
  was $16^4$.}
\label{ICorrb11}}

In all the cases, the numerical results obtained with the geometric and
heat-bath algorithms essentially agree, except for the correlations of the
imaginary part of the plaquettes in the Coulomb phase (Fig. \ref{ICorrb11}),
where a clear discrepancy for distances larger or equal than $4$ is
observed. Again in this case the reason for this discrepancy is related, as
for the Wilson loop results previously discussed, to the formation of
extremely long-lived metastable states \cite{monos} in the heat-bath
simulations, which seem to be absent in the geometric algorithm.  Indeed we
have verified, with simulations in $12^4$ lattices, that when we start the
heat-bath runs from a cold configuration, the disagreement on
the correlations of the imaginary part of the plaquettes in Coulomb phase at
large distances basically disappear. There are still small discrepancies in this case, but they
can be reasonably attributed to the difference in finite volume terms between
the two algorithms.

To compare computational costs we define a figure of merit, which is
the product of the squared error times the cpu time. We expect the
error to vanish like \mbox{$1/\sqrt{N_{\text {Monte Carlo}}}$}, and
therefore the quantity defined above should tend asymptotically to a
constant. We show the value of this quantity for several observables
in both phases and for both algorithms in Fig. \ref{Err}.
\FIGURE[float]{
\resizebox{12 cm}{!}{\includegraphics[angle=-90]{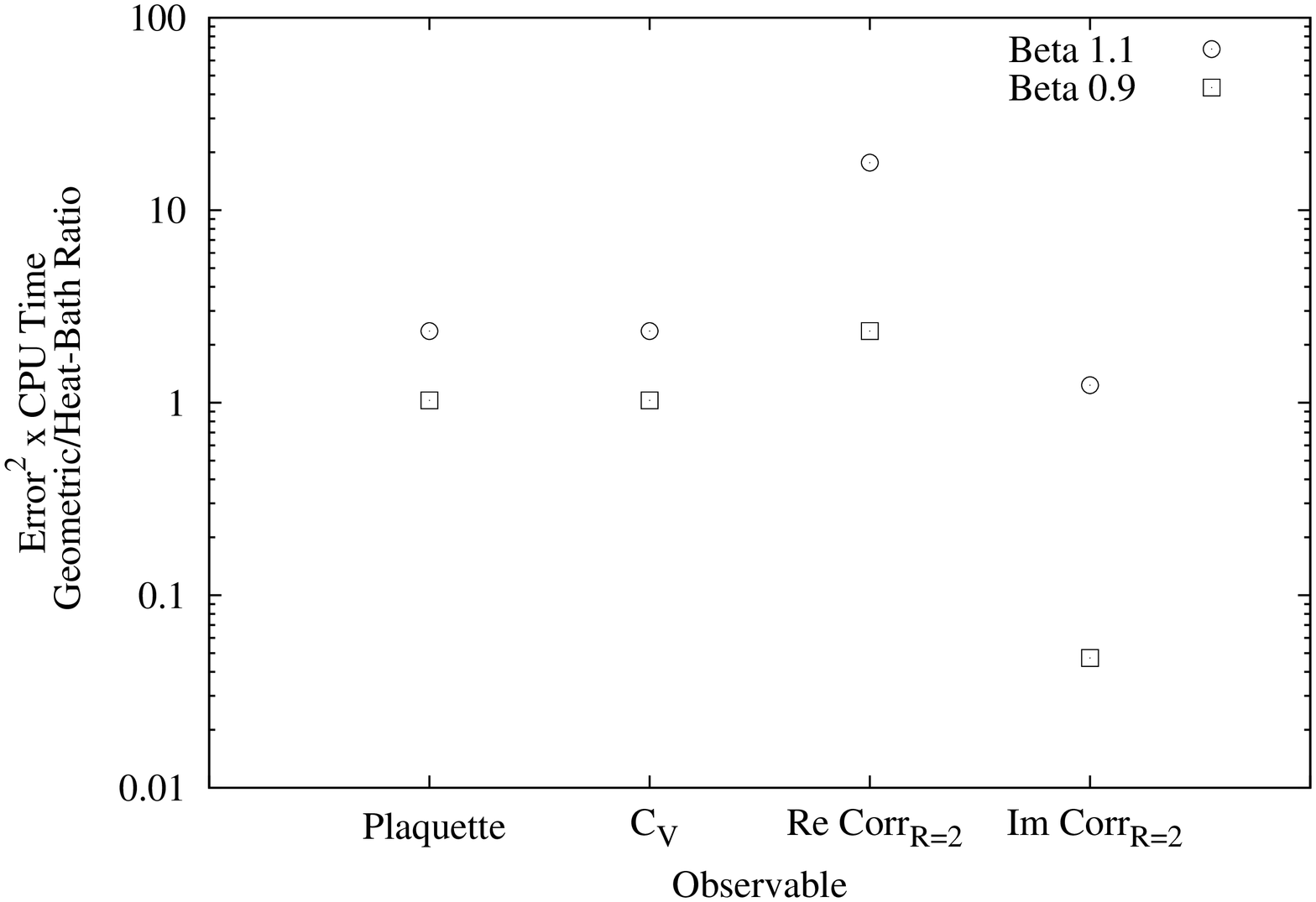}}
\caption{Ratios of figures of merit for different observables between
  geometric and heat-bath algorithms. Re Corr$_{R=2}$ and Im Corr$_{R=2}$ stand
  for Real and Imaginary plaquette--plaquette correlations at distance $2$.}
\label{Err}}
We can see that the performance of both algorithms is quite comparable. The
differences that are seen could conceivably change if one were to optimize the
specific implementations, but none is obviously much more efficient than the
other for the models studied.

In particular, for the plaquette observable and the specific heat, both
algorithms have a similar figure of merit. From our point of view, the
differences are not quite significant, and could change with careful
optimizations. The real plaquette-plaquette correlation is quite another
story, for the differences become significative in the Coulomb phase (a factor
$\approx 20$), but they do not become worse as $\beta$ increases, as we test
in a $12^4$ simulation at $\beta = 3.0$.

On the other hand, the geometric algorithm seems to perform much
better for the imaginary plaquette-plaquette correlation in the
confining phase, whereas in the Coulomb phase all the advantage
vanishes. Again, our $12^4$ computation at $\beta = 3.0$ reveals that
the ratio slowly decreases as $\beta$ increases (being $\approx 0.8$
at $\beta=3.0$).

Of course, this analysis assumes that both algorithms have no ergodicity
problems. We must be careful to start from a cold configuration when running
the heat-bath simulations in the Coulomb phase, in order to avoid metastable
states which could spoil the reliability of the simulation.

\section{The three dimensional Ising gauge model}

Let us finally come to the point of critical slowing down: This is a major
issue, as any improvement in this field can be of paramount importance in term
of the cost of large scale simulations of (statistical) systems at a critical
point. Beating critical slowing down is one of the main motivations in the
development of new Monte Carlo algorithms.

Typically what is found in Monte Carlo simulations of system both in
statistical physics and gauge theories is that the autocorrelation time $\tau$
diverges as we approach a critical point, usually as a power of the spatial
correlation length: $\tau \sim \xi^z$, where $\xi$ is the correlation length
and $z$ is a dynamical critical exponent. Usual local algorithms have values
of $z$ around 2, making it very inefficient to simulate close to the critical
point. For spin systems there are well known cluster algorithms with much
smaller $z$. Previously published results \cite{Sokal} on an algorithm similar
to ours, but applied to a non-gauge model, have claimed a similarly smaller
value for $z$. Having also this motivation in mind, we have investigated the
autocorrelation properties of our numerical scheme on the critical point of a
system that undergoes a second order phase transition (with diverging
correlation length). Our model of choice has been the three dimensional
Ising-gauge model. We have performed extensive simulations in the critical
zone of this model for several values of the lattice size (and hence
correlation length), using both the geometric algorithm and the standard Monte
Carlo approach, the latter known to have a lower bound for the autocorrelation
exponent $z$ equal to $2$, a value typical of all local algorithms. For
lattices up to $L = 24$ we have in all cases more than $5 \times 10^5$
Monte-Carlo iterations, which increase to more than $1 \times 10^6$ for $L =
32, 48$, and to more than $4 \times 10^6$ iterations for the largest lattice
$L = 64$.

For an observable $O$ we define the autocorrelation function $\rho(t)$
as
\be
\rho(t) = \frac{\left\langle\left(O(i) - O_A\right)
\left(O(i+t) - O_B\right)\right\rangle}{\sqrt{\sigma_A^2 \sigma_B^2}} 
\ee
where 
$O_A = \left\langle O(i)\right\rangle$, $O_B = \left\langle
  O(i+t)\right\rangle$, $\sigma_A^2 = \left\langle\left(O(i) -
    O_A\right)^2\right\rangle$, $\sigma_B^2 = \left\langle\left(O(i+t)
    - O_B\right)^2\right\rangle$,
and $\left\langle \right\rangle$ denotes average over i.  We then
define the integrated autocorrelation time by
\be
\tau = \rho(0) + 2\sum_{t=1}^N \rho(t)\frac{N - t}{N}
\ee
where $N$ is fixed, but with $N< 3\tau$ and $N < 10\%$ of the total sample. In
Fig. \ref{tau} we report the results for the integrated autocorrelation time
of the plaquette versus lattice size in logarithmic scale for both algorithms.
\FIGURE[float]{
\resizebox{12 cm}{!}{\includegraphics[angle=-90]{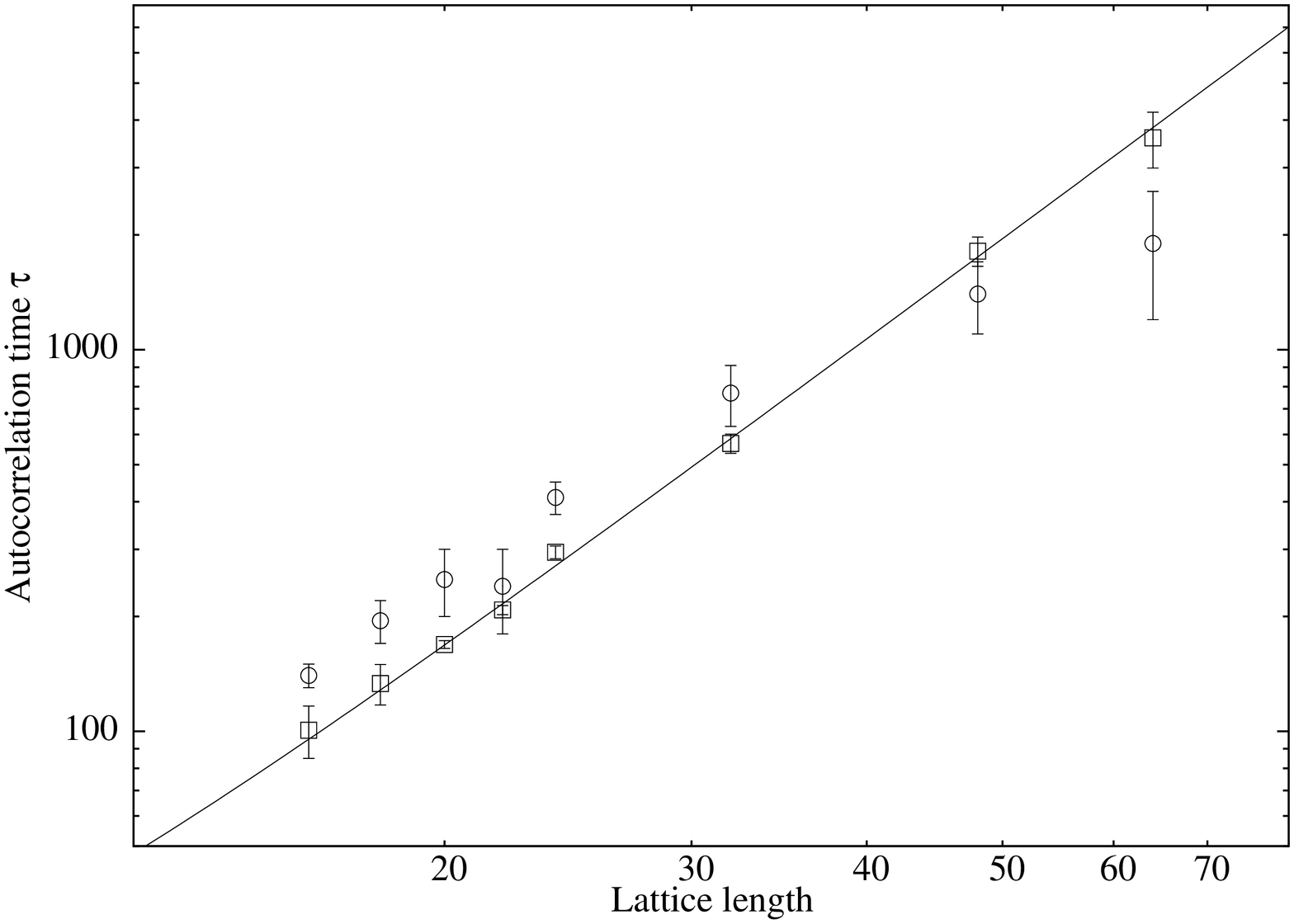}}
\caption{Autocorrelation times at the critical point (of each
  algorithm) versus lattice length; boxes stand for standard algorithm
  results, with a linear fit to guide the eye, while circles represent
  the results of the geometric algorithm. The errors were obtained by
  a jack-knife procedure.}
\label{tau}}
The results of our simulations hint to a different asymptotic
behaviour of the autocorrelation time, although with our present data
we cannot obtain a conclusive result. The results for the heat-bath
algorithm seem to fall nicely on a straight line, which would
correspond to a simple exponential dependence of $\tau$ on $L$, with
$z = 2.67\pm0.08$, but the geometric algorithm presents a more
complicated behaviour, as well as larger errors. There are signs
that the asymptotic behaviour might be better than for the heat-bath,
but much more extensive simulations, outside the scope of this work,
would be needed to get a definite value for $z$.

\section{Conclusions and Outlook}

Our main motivation is the sign problem, particularly in QCD at finite
chemical potential, but which also appears in other systems of
interest, for example when a $\theta$ vacuum term is present. New
ideas are clearly needed in order to make significant advances in this
problem, and one possibility is the development of new simulation
algorithms that might circumvent the difficulties of conventional
approaches. 

We have developed a geometric algorithm, based on the strong coupling
expansion of the partition function, which can be applied to abelian
pure gauge models. We have checked in the $U(1)$ model in 3 and 4
dimensions that the algorithm can be implemented efficiently, and is
comparable with a standard heat-bath algorithm for those models. It
seems however that the geometric algorithm does not suffer lack of
ergodicity due to the presence of vortices, as can be the case for
heat-bath, depending on the starting point.

We have also studied the algorithm in the 3 dimensional Ising gauge
model at the critical point, where we have seen hints that the
asymptotic behaviour of the geometric algorithm may be better than
standard heat-bath. This would be very interesting, because in
contrast to spin systems, where there exists cluster algorithms that
can greatly reduce critical slowing-down, to our knowledge no similar
algorithm is known for gauge systems. Our results are however not
enough to establish this, and much more extensive simulations should
be done to clarify this point.

The algorithm can be extended to include fermions, and this
constitutes work in progress. In this case there is a sign problem,
and one question we want to answer is whether such problem  is
severe or mild.

\section*{Acknowledgments}

This work has been partially supported by
an INFN-MEC collaboration, CICYT (grant FPA2006-02315)
and DGIID-DGA (grant2007-E24/2).
E. Follana is supported by Ministerio the Ciencia e Innovaci\'on
through the Ram\'on y Cajal program.

\end{document}